\input amstex
\documentstyle{amsppt}
\document
\magnification=1200
\NoBlackBoxes
\nologo
\vsize18cm

\centerline{\bf CLASSICAL COMPUTING, QUANTUM COMPUTING,}

\medskip

\centerline{\bf AND SHOR'S FACTORING ALGORITHM\footnote{Talk at 
the Bourbaki Seminar, June 1999.}}

\medskip

\centerline{\bf Yu.~I.~Manin}

\medskip

\centerline{\it Max--Planck--Institut f\"ur Mathematik, Bonn, Germany}

\bigskip

\centerline{\bf 0. Why quantum computing?}

\medskip

Information processing (computing) is the dynamical evolution of
a highly organized physical system produced by technology
(computer) or nature (brain). The initial state of this
system is (determined by) its input; its final state is the output.
Physics describes nature in two complementary modes:
classical and quantum. Up to the nineties, the basic mathematical
models of computing, Turing machines, were 
classical objects, although the first suggestions for studying
quantum models date back at least to 1980. 

\smallskip

Roughly speaking, the motivation to study quantum computing
comes from several sources: physics and technology, cognitive science,
and mathematics. We will briefly discuss
them in turn.

\smallskip

(i) Physically, the quantum mode of description is more fundamental
than the classical one. In the seventies and  eighties it was
remarked that, because of the superposition principle,
it is computationally unfeasible to simulate
quantum processes on classical computers ([Po], [Fe1]). Roughly
speaking, quantizing a classical system with
$N$ states we obtain a quantum system whose
state space is an $(N-1)$--dimensional complex projective space
whose volume grows {\it exponentially} with $N.$
One can argue that the main preoccupation of quantum chemistry
is the struggle with resulting difficulties.
Reversing this argument, one might expect that
quantum computers, if they can be built at all,
will be considerably more powerful than classical ones ([Fe1], [Ma2]).

\smallskip

Progress in the microfabrication techniques
of modern computers has already led us to
the level where quantum noise becomes an essential
hindrance to the error--free functioning of microchips.
It is only logical to start exploiting the
essential quantum mechanical behavior of small objects in devising
computers, instead of neutralizing it.

\smallskip

(ii) As another motivation, one can invoke highly speculative, but intriguing, conjectures that our brain is in fact
a quantum computer.
For example, the recent progress in writing
efficient chess playing software (Deep Blue) shows that to simulate
the world championship level using only classical algorithms,
one has to be able to analyze about $10^6$ positions/sec
and use about $10^{10}$ memory bytes. Since the
characteristic time of  neuronal processing is about
$10^{-3}$ sec, it is very difficult to explain how the
classical brain could possibly do the job and play chess
as successfully as Kasparov does. A less spectacular, but not less resource
consuming task, is speech generation and 
perception, which is routinely done by billions
of human brains,  but still presents a formidable challenge
for modern computers using classical algorithms.

\smallskip

Computational complexity of cognitive tasks has several
sources: basic variables can be fields; a restricted amount of
small blocks can combine into exponentially growing trees
of alternatives; databases of incompressible information
have to be stored and searched.

\smallskip

Two paradigms have been developed to cope with these
difficulties: logic--like languages and combinatorial
algorithms, and  statistical matching of
observed data to an unobserved model (see D.~Mumford's
paper [Mu] for a lucid discussion of the second paradigm.)

\smallskip

In many cases, the second strategy efficiently supports
an acceptable performance, but usually cannot achieve
excellency of the Deep Blue level. Both paradigms require
huge computational resources, and it is not clear,
how they can be organized, unless hardware
allows massive parallel computing.

\smallskip

The idea of 
``quantum parallelism'' (see sec. 2 below) is
an appealing theoretical alternative. However, it is not at all clear
that it can be made compatible with
the available experimental evidence, which depicts
the central nervous system as a distinctly
classical device. 

\smallskip

The following way out might be worth exploring.
The implementation of efficient quantum algorithms 
which have been studied so far
can be  provided by one, or several, quantum chips
(registers) controlled
by a classical computer. A very considerable part of
the overall computing job, besides controlling quantum chips,
is also assigned to the classical computer.
Analyzing a physical device of such architecture,
we would have  direct access to its
classical component (electrical or neuronal network),
whereas locating its quantum components might constitute
a considerable challenge. For example, quantum chips
in the brain might be represented by macromolecules of the type
that were considered in some theoretical models for 
high temperature superconductivity.  

\smallskip

The difficulties are
seemingly increased by the fact that
quantum measurements produce non--deterministic outcomes.
Actually, one could try to use this to one's advantage, 
because there exist 
situations where we can distinguish  the quantum randomness
from  the classical one by analyzing the probability
distributions and using the Bell--type inequalities.
With hindsight,
one recognizes in Bell's setup the first example
of the game--like situation where quantum players
can behave demonstrably more efficiently that
the classical ones (cf. the description of this setup
in [Ts], pp. 52--54).

\smallskip

It would be extremely interesting 
to devise an experimental setting purporting to show that
some  fragments of the central
nervous system relevant for information processing  
can in fact be in a quantum
superposition of classical states.

\smallskip

(iii) Finally, we turn to mathematics. One can argue that
nowadays one does not even need  additional motivation,
given the predominant mood prescribing the
quantization of ``everything that moves''. Quantum groups, quantum
cohomology, quantum invariants of knots etc come to mind.
This actually seemed to be the primary motivation
before 1994, when P.~Shor ([Sh]) devised the first quantum algorithm
showing that prime factorization can be done on quantum
computers in polynomial time, that is, considerably faster than
by any known classical algorithm. (P.~Shor's work was
inspired by the earlier work [Si] of D.~Simon). Shor's
paper gave a new 
boost to the subject. Another beautiful result due to L.~Grover ([Gro])
is that a quantum search among $N$ objects can be
done in $c\sqrt N$ steps. A.~Kitaev [Ki1] devised new
quantum algorithms for computing stabilizers
of abelian group actions; his work was
preceded by that of D.~Boneh and R.~Lipton [BoL],
who treated the more general problem
by a modification of Shor's method (cf. also [Gri]). 
At least as important as the results
themselves, are the tools invented by Shor, Grover, and Kitaev.

\smallskip

Shor's work is the central subject of this lecture. It is
explained in sec. 4. This explanation follows the
discussion of the general principles of quantum computing
and massive quantum parallelism in sec. 2, and of four
quantum subroutines, including Grover's searching algorithm,
in sec. 3. The second of these subroutines involving
quantum computations of classical computable functions
shows how to cope with the basic issue of quantum reversibility
vs classical irreversibility.
For more on this, see [Ben1] and [Ben2]. The opening sec. 1 contains
a brief report on the classical theory of computability.
I made some effort to express certain notions
of computer science, including P/NP, in the 
language of mainstream mathematics. The last section 5
discusses Kolmogorov complexity in the context
of classical and quantum computations.

\smallskip

Last, but not least, the hardware for quantum
computing does not exist as yet: see 3.3 below for
a brief discussion of the first attempts to
engineer it. The quantum 
algorithms invented and studied up to now will stimulate
the search of technological implementation which -- if
successful -- will certainly correct our present understanding
of quantum computing and quantum complexity.

\medskip

{\it Acknowledgements.} I am grateful to Alesha Kitaev,
David Mumford, and Dimitri Manin for their interest
and remarks on the earlier version of this report.
Many of their suggestions are incorporated in the text.

\bigskip

\centerline{\bf 1. Classical theory of computation}

\medskip

{\bf 1.1. Constructive universe.} In this section I deal only with
deterministic computations, which can be modelled by 
classical discrete time dynamical systems and subsequently
quantized. 

\smallskip

Alan Turing undertook
the microscopic analysis of the intuitive idea of algorithmic computation.
In a sense, he found its genetic code.
The atom of information is one bit, the
atomary operators can be chosen to act upon one/two bits and to produce
the outputs of the same small size. Finally, the sequence
of operations is strictly determined by the local environment
of bounded size, again several bits.

\smallskip

For a change, I proceed in the reverse direction,
and start this section with a 
presentation of the macrocosm of the classical theory of computation.
Categorical language is appropriate to this end.

\smallskip

Let $\Cal{C}$ be a category whose objects are countable or 
finite sets $U$. Elements $x$ of these sets will generally 
be finite sets with additional structure. Without
waiting for  all the necessary
axioms to be introduced, we will call $x\in U$ {\it a constructive
object of type $U$} (an integer, a finite graph, a word in
a given alphabet, a Boolean expression, an instance of a mass
problem $\dots$) The set $U$ itself will be
called  {\it the constructive world} of objects of fixed type, and
$\Cal{C}$ {\it the constructive universe.} The category $\Cal{C},$
which will be made more concrete below, will 
contain all finite products and finite unions of its objects,
and also finite sets $U$ of all cardinalities.

\smallskip

Morphisms $U\to V$ in $\Cal{C}$ are certain partial
maps of the underlying sets. More precisely, such a morphism
is a pair $(D(f),f)$ where $D(f)\subset U$ and
$f:\,D(f)\to V$ is a set--theoretic map. Composition is
defined by 
$$
(D(g),g)\circ (D(f),f)=(g^{-1}D(f),g\circ f).
$$
We will omit $D(f)$ when it does not lead to a confusion.

\smallskip

The morphisms $f$ that we will be considering are
{\it (semi)computable functions} $U\to V.$ An intuitive
meaning of this notion, which has a very strong
heuristic potential, can be explained as follows:
there should exist an algorithm $\varphi$
such that if one takes as  input the constructive
object $u\in U,$ one of the three alternatives holds:

\smallskip

(i) $u\in D(f),$ $\varphi$ produces in a finite number of steps
the output $f(u)\in V.$

(ii) $u\notin D(f)$, $\varphi$ produces in a finite number of steps
the standard output meaning NO.

(iii) $u\notin D(f)$, $\varphi$ works for an infinitely long time
without producing any output.

\smallskip

The necessity of including the alternative (iii) in the
definition of (semi--)computa\-bility was an important and non--trivial
discovery of the classical theory. The set of all morphisms
$U\to V$ is denoted $\Cal{C}(U,V).$

\smallskip

The sets of the form $D(f)\subset U$ are called
{\it enumerable subsets of} $U.$ If both $E\subset U$ and
$U\setminus E$ are enumerable, $E$ is called {\it decidable.}

\smallskip

The classical computation theory makes all of this more precise
in the following way.
 
\medskip

\proclaim{\quad 1.2. Definition} A category $\Cal{C}$ as above is called
a constructive universe if it contains 
the constructive world $\bold{N}$ of all integers $\ge 1,$ finite sets
$\emptyset,\,\{1\},\dots $, $\{1,\dots ,n\},
\dots$
and satisfies the following conditions
(a)--(d).

\smallskip

(a) $\Cal{C}(\bold{N}, \bold{N})$ is defined
as the set of all partially recursive functions
(see e.g. [Ma1],
Chapter V, or [Sa]).

\smallskip
    
(b) Any infinite object of $C$ is isomorphic to $\bold{N}.$

\smallskip

(c) If $U$ is finite, $\Cal{C}(U,V)$ consists of all
partial maps $U\to V.$ If $V$ is finite,
$\Cal{C}(U,V)$ consists of such $f$ that inverse image of any element
of $V$ is enumerable.

\endproclaim

\medskip

Before stating the last condition (d), we make some comments.

\smallskip

Statement (b) is a part of the famous Church Thesis.
Any isomorphism (computable bijection) 
$\bold{N}\to U$ in $\Cal{C}$ is called a {\it numbering.}
Thus, two different numberings of the same constructive
world differ by a recursive permutation of $\bold{N}.$
We will call such numberings equivalent ones.
Notice that because of (c) two finite constructive worlds 
are isomorphic iff they have the same cardinality,
and the automorphism group of any finite $U$
consists of all permutations of $U.$

\smallskip

As a matter of principle, we always consider $\Cal{C}$ as an open category,
and at any moment allow ourselves to add to it new constructive worlds.
If some infinite $V$ is added to $\Cal{C}$, it must come
together with a class of equivalent numberings.
Thus, any finite union of constructive worlds
can be naturally turned into the constructive world,
so that the embeddings become computable morphisms,
and their images are decidable.
As another example, the world $\bold{N}^{*}$  of finite sequences 
of numbers from $\bold{N}$ (``words in alphabet
$\bold{N}$'') is endowed with G\"odel's numbering
$$
(n_1,n_2,\dots ,n_k, \dots )\mapsto 2^{n_1-1}3^{n_2-1}\dots p_k^{n_k-1}\dots
\eqno(1)
$$
where $p_k$ is the $k$--th prime number. Hence we may
assume that $\Cal{C}$ is closed with respect to the
construction $U\mapsto U^*.$ All natural functions, such as
length of the word $U^*\to\bold{N}$, or the $i$--th letter
of the word $U^*\to U$ are computable.

\smallskip

Similarly, $\Cal{C}$ can be made closed with respect to the finite
direct products by using the (inverse) numbering of $\bold{N}^2$:
$$
(m,n)\mapsto m+\frac{1}{2}\,(m+n-1)(m+n-2).
\eqno(2)
$$
Projections, diagonal maps, fiber maps $V\to U\times V,
v\mapsto (u_0,v)$ are all computable.

\smallskip

Decidable subsets of constructive worlds are again constructive.

\smallskip

Church Thesis is often invoked as a substitute for an explicit
construction of a numbering, and it 
says that {\it the category $\Cal{C}$ is defined uniquely
up to equivalence.}

\smallskip

We now turn to the computability properties of the sets of
morphisms $\Cal{C}(U,V).$ Again, it is a matter of principle that 
$\Cal{C}(U,V)$ itself
{\it is not a constructive world if $U$ is infinite.}
To describe the situation axiomatically, consider first 
any diagram 
$$
\roman{ev}:\, P\times U\to V
\eqno(3)
$$ 
in $\Cal{C}.$
It defines a partial map $P\to\Cal{C}(U,V),\, p\mapsto\overline{p},$
where $\overline{p}(u):=\roman{ev}\,(p,u).$ We will say that the constructive world
$P=P(U,V)$ together with the evaluation map $\roman{ev}$ is {\it a programming method} 
(for computing some maps $U\to V$). It is called {\it universal,}
if the following two conditions are satisfied. First,
the map $P\to\Cal{C}(U,V)$ must be surjective.
Second, for any programming method $Q=Q(U,V)$ with the same
source $U$ and target $V,$ 
$\Cal{C}(Q,P)$ contains translation morphisms
$$
\roman{trans}:\,Q(U,V)\to P(U,V)
\eqno(4)
$$
which are, by definition, everywhere defined, computable maps
$Q\to P$ such that if $q\mapsto p,$ then $\overline{q}=\overline{p}.$

\smallskip

We now complete the Definition 1.2 by adding the last axiom
forming part of the Church Thesis:

\medskip

{\it (d) For every two constructive worlds $U,V,$ there exist
universal programming methods.}

\medskip

The standard examples of $P$ for $U=V=\bold{N}$ are
(formalized descriptions of) Turing machines,
or recursive functions. 

\smallskip

From (d) it follows that the composition of morphisms 
can be lifted to a
computable function on the level of programming
methods. To be more precise, if $Q$ (resp. $P$)
is a programming method for $U,V$ (resp. $V,W$),
and $R$ is a universal programming method for $U,W,$
there exist computable composition maps
$$
\roman{comp}:\, P(V,W)\times Q(U,V)\to R(U,W),\, (p,q)\mapsto r
\eqno(5)
$$ 
such that $\overline{r}=\overline{p}\circ\overline{q}.$ 

\smallskip

Concrete $P(U,V)$ are furnished by the choice of
what is called the ``model of computations''
in computer science. This last notion comes with a detailed 
description not only of programs but also of
all steps of the computational
process. At this stage the models of kinematics
and dynamics of the process first emerge,
and the discussion of quantization can start.

\smallskip

A formalized description of the first
$n$ steps will be called {\it a history of
computation} or, for short, {\it a protocol}
(of length $n$.) For a fixed model,
protocols (of all lenghts) form a constructive world as well.
We will give two formalized versions of this
notion, for functions
with infinite and finite domains
respectively. The first will be well suited for the discussion of  
polynomial time computability, the second is the base
for quantum computing.

\medskip

{\bf 1.3. Models of computations I: normal models.}
Let $U$ be an infinite constructive world.
In this subsection we will be considering partial functions
$U\to U.$ The more general case $U\to V$ can be reduced to this
one by working with $U\coprod V.$ 

\smallskip

{\it A normal model of computations}
is the structure $(P,U,I,F,s,)$ consisting of four sets
and a map:
$$
I\subset U,\  F\subset P\times U,\ s:\,P\times U\to P\times U\, .
\eqno(6)
$$
Here $s$ is an everywhere defined function
such that $s(p,u)=(p,s_p(u))$ for
any $(p,u)\in P\times U.$ Intuitively, $p$ is a program,
$u$ is a configuration of the deterministic discrete time
computing device,
and $s_p(u)$ is the new configuration obtained from $u$ after one
unit of time (clock tick). Two additional subsets $I\subset U$ (initial
configurations, or inputs) and $F\subset P\times U$ (final configurations)
must be given,  
such that if $(p,u)\in F,$ then $s(p,u)=(p,u)$ i.e. $u$
is a fixed point of $s_p.$ 

\smallskip

In this setting, we denote by $f_p$ the partial function
$f_p:\,I\to U$ such that we have 
$$
u\in D(f_p)\ \roman{and}\ f_p(u)=v\
\roman{iff\ for\ some}\ n\ge 0,\ (p,s_p^n(u))\in F\ \roman{and}\ s_p^n(u)=v.
\eqno(7)
$$ 
The minimal such $n$ will be called the {\it time}
(number of clock ticks) needed to calculate $f_p(u)$ using the program $p.$

\smallskip

Any finite sequence
$$
(p, u, s_p(u), \dots , s_p^m(u)), \ u\in I,
\eqno(8)
$$
will be called {\it a protocol of computation of length $m.$}

\smallskip

We now add the constructivity conditions.

\smallskip

We require $P,U$ to be constructive worlds, $s$ computable.
In addition, we assume that $I,F$ are decidable subsets of
$U,$ $P\times U$ respectively. Then $f_p$ are computable, 
and protocols of given length, (resp. of arbitrary length, resp.
or those stopping at $F$), 
form constructive worlds. If we denote by $Q$ the world
of protocols stopping at $F$ and by $\roman{ev}:\,Q\times U\to U$
the map $(p,u)\mapsto s_p^{\roman{max}}(u)$, we get a programming
method. 

\smallskip

Such a model is called {\it universal,} if the respective
programming method is universal.

\smallskip

The notion of normal model of computations generalizes
both normal algorithms and Turing machines.
For their common treatment see e.g. [Sa], Chapter 4.
In broad terms, $p\in P$ is the list of Markov
substitutions, or the table defining the operation
of a Turing machine. The remaining worlds $U,I,F$ consist
of various words over the working alphabet. 

\smallskip

\smallskip

{\bf 1.3.1. Claim.} {\it For any $U$,
universal normal models of computations exist, and can be
effectively constructed.}

\smallskip

For $U=\bold{N},$ this follows from the existence of universal Turing machines,
and generally, from the Church Thesis.
It is well known that the universal machine for calculating
functions of $k$ arguments is obtained by
taking an appropriate function of $k+1$ arguments
and making the first argument the variable part of the
program. Hence $P,$ in this case, consists of pairs
$(q,m),$ where $q$ is the fixed program of the
$(k+1)$--variable universal function (hardware) and $m$ is
a word written on the tape (software).  
 
\medskip

{\bf 1.4. Models of computations II: Boolean circuits.}
Boolean circuits are classical models
of computation well suited for studying maps between the 
finite sets whose elements are encoded
by sequences of 0's and 1's. 

\smallskip

Consider the Boolean algebra $\bold{B}$ generated over $\bold{F}_2$
by a countable sequence of independent variables, say $x_1,x_2,x_3,
\dots$ This is the quotient algebra of $\bold{F}_2[x_1,x_2,
\dots]$ with respect to the relations $x_i^2=x_i.$ Each Boolean
polynomial determines a function on 
$\oplus_{i=1}^{\infty}\bold{F}_2$ with values in $\bold{F}_2=\{0,1\}.$

\smallskip

We start with the following simple fact.

\smallskip

{\bf 1.4.1. Claim.} {\it Any map $f:\,\bold{F}_2^m\to \bold{F}_2^n$
can be represented by a unique vector of Boolean polynomials.}

\smallskip

{\bf Proof.} It suffices to consider the case $n=1.$ Then
$f$ is represented by
$$
F(x_1,\dots ,x_n):=\sum_{y=(y_i)\in\bold{F}_2^m}f(y)\,
\prod_i (x_i+y_i+1)
\eqno(9)
$$
because the  product in (9) is the delta function in $x$
supported by $y.$ Moreover, the spaces of maps and
of Boolean polynomials have the common dimension $2^m$
over $\bold{F}_2.$

\smallskip

Now we can calculate any vector of Boolean polynomials iterating
operations from a small finite list, which is chosen and fixed,
e.g. $\Cal{B}:= \{x,\,1,\, x+y,\,xy,\,(x,x)\}.$
Such operators are called {\it classical gates.}
A sequence of such operators, together with indication
of their arguments from the previously computed
bits, is called {\it a Boolean circuit.} The number of steps
in such a circuit is considered as (a measure of) 
the time of computation.

\smallskip

When the relevant finite sets are not $\bold{F}_2^m$
and perhaps have a wrong cardinality, we encode
their elements by finite sequences of bits and consider
the restriction of the Boolean polynomial to
the relevant subset.

\smallskip

As above, a protocol of computation in this model can be represented as
the finite table consisting of rows (generally of variable length)
which accommodate sequences of 0's and 1's.
The initial line of the table is the input. 
Each subsequent line must be obtainable from
the previous one by the  application of one  
the basic functions in $\Cal{B}$ to the sequence of neighboring bits
(the remaining bits are copied unchanged). The last
line is the output. The exact location of the 
bits which are changed in each row and the nature of change
must be a part of the protocol.

\smallskip

Physically, one can implement the rows as the different
registers of the memory, or else as the consecutive
states of the same register (then we have to make a prescription for
how to cope with the variable length, e.g. using blank symbols).

\medskip

{\bf 1.4.2. Turing machines vs Boolean circuits.} 
Any protocol of the Turing computation of a function can be treated
as such a protocol of an appropriate Boolean circuit, 
and in this case we have only
one register (the initial part of the tape)
whose states are consecutively changed by the head/processor.
We will still use the term ``gate'' in this context. 

\smallskip

A computable function $f$ with infinite domain
is the limit of a sequence of functions $f_i$ between
finite sets whose graphs extend each other.
A Turing program for $f$
furnishes a computable sequence of Boolean circuits,
which compute all $f_i$ in turn. Such a sequence
is sometimes called {\it uniform}.

\medskip

{\bf 1.5. Size, complexity, and polynomial time computability.}
The quantitative theory of computational models deals simultaneously with
the space and time dimensions of protocols. The preceding
subsection focused on time, here we introduce space.
For Boolean (and Turing machine) protocols
this is easy: the length of each row of the protocol
is the space required at that moment (plus several more bits
for specifying the next gate). The maximum of these lengths
is the total space required.

\smallskip

The case of normal models and infinite
constructive worlds is more interesting.

\smallskip

Generally we will call {\it a size function} $U\to \bold{N}:\, u\to |u|$
any function such that for every $B\in\bold{N},$ there are
only finitely many objects with $|u|\le B.$
Thus the number
of bits $|n|=[\roman{log}_2n]+1$ and the identical function $\|n\|=n$
are both size functions. Using a numbering,
we can transfer them to any constructive world.
In these two examples, the number of constructive
objects of size $\le H$ grows as $\roman{exp}\,cH$, resp. $cH.$ 
Such a count in more general cases allows one to make a distinction
between {\it the bit size,} measuring the length of a description
of the object, and {\it the volume} of the object.

\smallskip

In most cases we require computability of size functions.
However, there are exceptions: for example, Kolmogorov
complexity is a non--computable size function with
very important properties: see below and sec. 5.

\smallskip

Given a size function (on all relevant worlds)
and a normal model of computations $\Cal{S}$, we can consider the following complexity problems.

\medskip

(A) {\it For a given morphism (computable map) $f:\,U\to V$,
estimate the smallest size $K_{\Cal{S}}(f)$ of the program $p$ such that
$f=f_p.$} 

\smallskip

Kolmogorov, Solomonoff and Chaitin proved that
there exists an {\it optimal} universal model of computations
$\Cal{U}$ such that, with $P=\bold{N}$ and the bit size function,
for any other model $\Cal{S}$
there exists a constant $c$ such that for any $f$
$$
K_{\Cal{U}}(f)\le K_{\Cal{S}}(f) + c.
$$
When $\Cal{U}$ is chosen, $K_{\Cal{U}}(f)$ is called
Kolmogorov's complexity of $f.$ With a different choice of
 $\Cal{U}$ we will
get the same complexity function up to $O(1)$--summand.

\smallskip

This complexity measure is highly non--trivial
(and especially interesting) for an one--element world $U$
and infinite $V.$
It measures then the size of the most
compressed description of a variable constructive object in $V.$
This complexity is quite ``objective'' being almost
independent of any arbitrary choices. Being uncomputable,
it cannot be directly used in  computer science.
However, it furnishes some basic restrictions on various
complexity measures, somewhat similar to those provided
by the conservation laws in physics.

\smallskip

On $\bold{N}$ we have $K_{\Cal{U}}(n) \le |n| +O(1)=\roman{log}_2\|n\|+O(1).$
The first inequality ``generically'' can be replaced by equality,  
but infinitely often $K_{\Cal{U}}(n)$ becomes
much smaller that $|n|.$

\smallskip

(B) {\it For a given morphism (recursive map) $f:\,U\to V$,
estimate the time needed to calculate $f(u), u\in D(f)$
using the program $p$ and compare the results for
different $p$ and different models of computations.}

\smallskip

(C) {\it The same for the function ``maximal size
of intermediate configurations in the protocol
of the computation of $f(u)$ using the program $p$'' (space, or memory).}

\medskip

In the last two problems, we have to compare functions
rather than numbers: time and space
depend on the size of input. Here a cruder polynomial scale appears naturally.
Let us show how this happens.

\smallskip

Fix a computational model $\Cal{S}$ with the transition function
$s$ computing functions $U\to U$, and choose a bit size function on $U$
satisfying the following crucial assumption:
\smallskip

$(\bullet )$ $|u|-c\le |s_p(u)|\le |u|+c$ {\it where the constant $c$ may depend
on $p$ but not on $u.$}

\smallskip

In this case we have $|s_p^m(u)|\le |u|+c_pm$: the required space
grows no more than linearly with time.

\smallskip

Let now $(\Cal{S}^{\prime},s^{\prime})$ be another model
such that $s_p=s_q^{\prime}$ for some $q.$ For example,
such $q$ always exists if $\Cal{S}^{\prime}$ is universal.
Assume that $s^{\prime}$ satisfies $(\bullet )$ as well,
and additionally

\smallskip

$(\bullet \bullet )$ {\it $s$ can be computed in the model
$\Cal{S}^{\prime}$ in time bounded by a polynomial $F$
in the size of input.}

\smallskip

This requirement is certainly satisfied for Turing
and Markov models, and is generally reasonable,
because an elementary step of an algorithm deserves
its name only if it is computationally tractable.

\smallskip
 
Then we can replace one application of $s_p$ to $s_p^{m}(u)$
by $\le F(|u|+cm)$ applications of $s_q^{\prime}.$
And if we needed $T(u)$ steps in order to calculate $f_p(u)$
using $\Cal{S},$ we will need no more than
$\le \sum_{m=1}^{T(u)} F(|u|+cm)$ steps to calculate
the same function using $\Cal{S}^{\prime}$ and $q.$
In a detailed model, there might be a small additional cost of 
merging two protocols. This is an example of the translation
morphism  (4) lifted to the worlds of protocols.

\smallskip

Thus, from $(\bullet )$ and $(\bullet \bullet )$
it follows that functions computable in polynomial
time by $\Cal{S}$ have the same property for all
reasonable models. Notice also that for such functions,
$|f(u)|\le G(|u|)$ for some polynomial $G$
and that the domain $D(f)$ of such a function
is decidable: if after $T(|u|)$ $s_p$--steps we are not in a final
state, then $u\notin D(f).$ 

\smallskip

Thus we can define the class $PF$
of functions, say, $\bold{N}^k\to\bold{N}$ 
computable in polynomial time
by using a fixed universal Turing machine and arguing as above
that this definition is model--independent. 

\smallskip

If we want to extend it to a constructive universe $\Cal{C}$
however, we will have to postulate additionally  that
any constructive world $U$ comes together with
a natural class of numberings which, together with their
inverses, are
computable in polynomial time. This seems to be a part of
the content of the {\it ``polynomial Church thesis''}
invoked by M.~Freedman in [Fr1]. If we take this
strengthening of the Church thesis for granted,
then we can define also the bit size
of an arbitrary constructive object as the bit size
of its number with respect to one of these numberings.
The quotient of two such size functions is
bounded from above and from zero.

\smallskip

Below we will
be considering only the universes $\Cal{C}$ and
worlds $U$ with these properties,
and $|u|$ will always denote one of the bit size
norms. G\"odel's numbering (2) for $\bold{N}\times\bold{N}$
shows that that such $\Cal{C}$ is still closed
with respect to finite products.  
(Notice however that the beautiful numbering (3)
of $\bold{N}^*$ using primes is not polynomial time
computable; it may be replaced by another one
which is in $PF$). 

\medskip

{\bf 1.6. $P/NP$ problem.} By definition, a subset $E\subset U$
{\it belongs to the class $P$} iff its characteristic
function $\chi_E$ (equal to 1 on $E$ and 0 outside)
belongs to the class $PF.$ Furthermore, $E\in U$ 
{\it belongs to the class $NP$} iff there exists
a subset $E^{\prime}\subset U\times V$ belonging to $P$
and a polynomial $G$ such that
$$
u\in E \iff \exists\, (u,v)\in E^{\prime}\ \roman{with}\
|v|\le G(|u|).
$$
Here $V$ is another world (which may coincide with
$U$). We will say that $E$ is obtained from $E^{\prime}$
by a {\it polynomially truncated projection.} 

\smallskip

The discussion above establishes in what sense
this definition is model independent. 

\smallskip

Clearly,
$P\subset NP.$ The inverse inclusion is highly
problematic. A naive algorithm calculating
$\chi_E$ from $\chi_{E^{\prime}}$ by searching for
$v$ with $|v|\le G(|u|)$ and $\chi_{E^{\prime}}(u,v)=1$ 
will take exponential time e.g. when there is no such
$v$ (because $|u|$ is a bit size function).
Of course, if one can treat all such $v$
in parallell, the required time will be polynomial.
Or else, if an oracle tells you that $u\in E$
and supplies an appropriate $v$, you can convince yourself
that this is indeed so in polynomial time,
by computing $\chi_{E{\prime}}(u,v)=1.$

\smallskip

Notice that the enumerable sets can be alternatively described
as projections of decidable ones, and that in this
context projection does create undecidable sets.
Nobody was able to translate the diagonalization
argument used to establish this to the $P/NP$ domain.
M.~Freedman ([Fr2]) suggested an exciting new approach to the problem
$P\ne NP$(?), based upon a modification of Gromov's strategy
for describing groups of polynomial growth.

\smallskip

It has long been known that this problem  can be reduced to checking
whether some very particular sets -- $NP$--complete ones --
belong to $P.$ The set $E\subset U$ is called {\it $NP$--complete}
if, for any other set $D\subset V, D\in NP,$ there exists
a function $f:\,V\to U, f\in PF,$ such that $D=f^{-1}(E),$
that is, $\chi_D(v)=\chi_E(f(v)).$
We will sketch the classical argument (due to
S.~Cooke, L.~Levin, R.~Karp) showing the existence
of $NP$--complete sets. In fact, the reasoning is
constructive: it furnishes a polynomially
computable map producing $f$ from the descriptions of $\chi_{E^{\prime}}$
and of the truncating polynomial $G.$

\smallskip

In order to describe one NP--complete problem,
we will define an infinite family of Boolean
polynomials $b_u$ indexed by the following data, constituting
objects $u$ of the constructive world $U$. One $u$ is a collection
$$
m\in \bold{N};\ \ (S_1,T_1),\dots ,(S_N,T_N), 
\eqno(10)
$$
where $S_i,\,T_i \subset \{1,\dots ,m\},$  and $b_u$ is defined as
$$
b_u(x_1,\dots ,x_m)=\prod_{i=1}^N\left(1+\prod_{k\in S_i}(1+x_k)
\prod_{j\in T_i}x_j\right).
\eqno(11)
$$
The size of (10) is by definition $|u|=mN.$

\smallskip

Put 
$$
E =\{u\in U\,|\,\exists v\in \bold{F}_2^m,\, b_u(v)=1\}.
$$
Using the language of Boolean truth values, one says that
$v$ {\it satisfies} $b_u$ if $b_u(v)=1$, and $E$ is called
the {\it satisfiability problem}, or $SAT.$ 

\medskip

{\bf 1.6.1. Claim.} $E\in NP.$

\smallskip

In fact, let 
$$
E^{\prime}= \{(u,v)\,|\,b_u(v)=1\}\subset U\times (\oplus_{i=1}^{\infty}
\bold{F}_2)\,.
\eqno(12)
$$
Clearly, $E$ is the full projection of $E^{\prime}.$
A contemplation will convince the reader that $E^{\prime}\in P.$  
In fact, we can calculate $b_u(v)$  performing
$O(Nm)$ Boolean multiplications and additions. The projection
to $E$ can be replaced by a polynomially truncated
projection, because we have to check only $v$ of size $|v|\le m.$

\medskip

{\bf 1.6.2. Claim.} {\it $E$ is $NP$--complete.}

\smallskip

In fact, let $D\in NP$, $D\subset A$ where $A$ is some universe.
Take a representation of $D$ as a polynomially truncated
projection of some set $D^{\prime}\subset A\times B, D^{\prime}\in P.$ 
Choose a normal, say Turing, model of computation and
consider the Turing protocols of computation of $\chi_{D^{\prime}}(a,b)$
with fixed $a$ and variable polynomially bounded $b.$ 
As we have explained above, for a given $a$, any such protocol
can be imagined as a table of a fixed
polynomially bounded size whose rows are the consecutive
states of the computation. In the ``microscopic''
description, the positions in this table can be filled only by $0$ or $1$.
In addition, each row is supplied by the specification
of the position and the inner state of the head/processor.
Some of the arrangements are valid protocols, others are not,
but the local nature of the Turing computation
allows one to produce a Boolean polynomial $b_u$ in appropriate
variables such that the valid protocols are recognized
by the fact that this polynomial takes value $1.$
For detailed explanations see e.g. [GaJ], sec. 2.6.
This defines the function $f$ reducing $D$ to $E.$
The construction is so direct that the polynomial
time computability of $f$ is straightforward.

\smallskip

Many natural problems are known to be $NP$--complete,
in particular 3--$SAT.$ It is defined as the subset of $SAT$
consisting of those $u$ for which $\roman{card}\,(S_i\cup T_i)=3$
for all $i$.

\medskip

{\bf 1.6.3. Remark.} Most of Boolean functions are not  computable
in polynomial time. Several versions of this statement can be
proved by simple counting. 

\smallskip

First of all, fix a finite basis $\Cal{B}$ of Boolean
operations as in 1.4.1, each acting upon $\le a$
bits. Then sequences of these operations of length $t$
generate $O((bn^a)^t)$ Boolean functions $\bold{F}_2^n\to\bold{F}_2^n$
where $b=\roman{card}\,\Cal{B}.$ On the other hand, the number
of all functions $2^{n2^n}$ grows as a double exponential of $n$
and for large $n$ cannot be obtained in time $t$
polynomially bounded in $n.$

\smallskip

The same conclusion holds if we consider not all functions
but only permutations: Stirling's formula for
$\roman{card}\,S_{2^n}=2^n!$ involves a double exponential.

\smallskip

Here is one more variation of this problem:
define the time complexity of a conjugacy class
in $S_{2^n}$ as the minimal number of steps needed
to calculate some permutation in this class.
This notion arises if we are interested
in calculating automorphisms of a finite universe
of cardinality $2^n$, which
is not supplied with a specific encoding
by binary words. Then it can happen that a judicious
choice of encoding will drastically
simplify the calculation of a given function.
However, for most functions we still will not be able
to achieve polynomial type computability,
because the asymptotical formula for the
number of conjugacy classes (partitions)
$$
p(2^n) \sim \frac{\roman{exp}\,(\pi\,\sqrt{\frac{2}{3}(2^n-\frac{1}{24}})}{
4\sqrt{3}(2^n-\frac{1}{24})}
$$ 
again displays the double exponential growth.

\bigskip

\centerline{\bf 2. Quantum parallelism}

\medskip

In this section we will discuss the basics:
how to use the superposition principle in order
to accelerate (certain) classical computations.

\smallskip

\proclaim{\quad 2.1. Description of the problem} Let
$N$ be a large number, $F:\,\{0,\dots ,N-1\}\to
\{0,\dots ,N-1\}$ a function such that the computation
of each particular value $F(x)$ is tractable,
that is, can be done in time polynomial
in $\roman{log}\,x.$ We want to compute
(to recognize) some property of the graph $(x,F(x)),$
for example:

\smallskip

(i) Find the least period $r$ of $F$, i.e.
the least residue $r\,\roman{mod}\,N$ such
that $F(x+r \,\roman{mod}\,N)=F(x)$ for all $x$
(the key step in the Factorization Problem.)

\smallskip

(ii) Find some $x$ such that $F(x)=1$ or establish 
that such $x$ does not exist (Search Problem.)
\endproclaim

\smallskip

As we already mentioned, the direct attack on such a problem consists
in compiling the complete list of pairs 
$(x,F(x))$ and then applying to it an
algorithm recognizing the property in question.
Such a strategy requires at least exponential time
(as a function of the bit size of $N$)
since already the length of the list is $N.$
Barring a theoretical breakthrough in understanding
such problems, (for example a proof that $P=NP$),
a practical response might be in exploiting
the possibility of parallel computing, i.e.
calculating simultaneously many -- or even all -- values of $F(x).$
This takes less time but uses (dis)proportionally more
hardware.

\smallskip

A remarkable suggestion due to D.~Deutsch (see [DeuJ], [Deu])
consists in using a quantum superposition of the
classical states $|x\rangle$  
as the replacement of the union of $N$ classical registers, each
in one of the initial states $|x\rangle$. To be more precise,
here is a mathematical model formulated as the definition.

\medskip

\proclaim{\quad 2.2. Quantum parallel processing: version I} Keeping
the notation above, assume moreover that $N=2^n$
and that $F$ is a bijective map (the set of all outputs
is a permutation of the set of all inputs). 

\smallskip

(i) The quantum space of inputs/outputs is the $2^n$--dimensional
complex Hilbert space $H_n$ with the orthonormal basis
$|x\rangle$, $0\le x \le N-1$. Vectors $|x\rangle$
are called classical states. 

\smallskip

(ii) The quantum version  of $F$ is the unique unitary operator
$U_F:\, H_n\to H_n$ such that $U_F|x\rangle=|F(x)\rangle .$

\smallskip

Quantum parallel computing of $F$ is (a physical realization of)
a system with the state space $H_n$ and the evolution operator
$U_F$.
\endproclaim

\smallskip

Naively speaking, if we apply $U_F$ to the initial state
which is a superposition of all classical states
with, say, equal amplitudes, we will get simultaneously
all classical values of $F$ (i.e. their superposition):
$$
U_F\,\left(\frac{1}{\sqrt{N}}\sum |x\rangle \right) 
= \frac{1}{\sqrt{N}}\sum |F(x)\rangle .
\eqno(14)
$$
We will now discuss various issues related to this definition,
before passing to its more realistic modification.

\medskip

(A) We put $N=2^n$ above because we are imagining the
respective classical system as
an $n$--bit register: cf. the discussion of Boolean
circuits. Every number $0\le x\le N-1$
is written in the binary notation $x=\sum_i\epsilon_i2^i$
and is identified with the pure (classical) state $|\epsilon_{n-1},\dots ,\epsilon_{0}\rangle$
where $\epsilon_i=0$ or $1$ is the state of the $i$--th register.
The quantum system $H_1$ is called {\it qubit.} We have
$H_n=H_1^{\otimes n},\, |\epsilon_{n-1},\dots ,\epsilon_{0}\rangle =
|\epsilon_{n-1}\rangle \otimes \dots \otimes |\epsilon_{0}\rangle . $

\smallskip
This conforms to the general principles of quantum mechanics.
The Hilbert space of the union of systems can be identified with
the tensor product of the Hilbert spaces of the subsystems.
Accordingly, decomposable vectors correspond to the states
of the compound for which one can say that the individual subsystems
are in definite states.

\medskip

(B) Pure quantum states, strictly speaking, are points of the
projective space $P(H_n)$ that is, complex lines in $H_n.$
Traditionally, one  considers instead vectors of norm one.
This leaves undetermined an overall phase factor $\roman{exp}\,i\varphi .$
If we have two state vectors, individual phase factors
have no objective meaning, but their quotient, that is the
difference of their phases, does have one. This difference 
can be measured by observing effects of interference.
This possibility is used
for implementing efficient quantum algorithms.

\medskip

(C) If a quantum system  $S$ is isolated, its dynamical evolution
is described by the unitary operator $U(t)=\roman{exp}\,iHt$
where $H$ is the Hamiltonian, $t$ is time. Therefore one option
for implementing $U_F$ physically is to design a device
for which $U_F$ would be a fixed time evolution operator.
However, this seemingly contradicts many deeply rooted notions
of the algorithm theory. For example, calculating $F(x)$
for different inputs $x$ takes different times, and it would
be highly artificial to try to equalize them already in the design.

\smallskip

Instead, one can try to implement $U_F$ as the result of a sequence
of brief interactions, carefully controlled by a classical
computer,  of  $S$ with environment
(say, laser pulses). Mathematically speaking, $U_F$
is represented as a product of some standard unitary operators 
$U_m\dots U_1$ each of which acts only on a small subset
(two, three) of classical bits. These operators are called
{\it  quantum gates.} 

\smallskip

The complexity
of the respective quantum computation is determined by its
length (the
number $m$ of the gates) and by the complexity
of each of them. The latter point is a subtle one: continuous
parameters, e.g. phase shifts, on which $U_i$ may depend,
makes the information content of each $U_i$
potentially infinite and leads to a suspicion that a
quantum computer will in fact perform an analog 
computation, only implemented in a fancy way.
A very interesting discussion in [Ts],
Lecture 9, convincingly refutes this viewpoint,
by displaying those features of quantum computation
which distinguish it from both analog and digital
classical  information processing. This discussion is based
on the technique of fault tolerant computing using
quantum codes for producing continuous variables
highly protected from external noise.

\medskip

(D) From the classical viewpoint, the requirement that $F$
must be a permutation looks highly restrictive (for instance,
in the search problem $F$ takes only two values). 
Physically, the reason for this requirement is that only such $F$ extend
to unitary operators (``quantum reversibility'').
The standard way out consists of introducing {\it two}
$n$--bit registers instead of one, for keeping the value
of the argument as well as that of the function.
More precisely, if $F(|x\rangle )$ is an arbitrary function,
we can replace it by the permutation $\widetilde{F}(|x,y\rangle):=
|x,F(x)\oplus y\rangle,$ where $\oplus$ is the Boolean (bitwise)
sum. This involves no more than a polynomial increase of the
classical complexity, and the restriction of $\widetilde{F}$
to $y=0$ produces the graph of $F$ which we need anyway
for the type of problems we are interested in.

\smallskip

In fact, in order to process a classical algorithm
(sequence of Boolean gates) for computing $F$ into the quantum one, we
replace each classical gate by the respective
reversible quantum gate, i.e. by the unitary
operator corresponding to it tensored
by the identical operator.  Besides two registers
for keeping $|x\rangle$ and $F(|x\rangle )$ 
this trick introduces as well extra qubits
in which we are not particularly interested. The corresponding
space and its content is sometimes referred to as ``scratchpad'',
``garbage'', etc. Besides ensuring reversibility,
additional space and garbage can be introduced as well
for considering functions $F:\,\{0,\dots ,N-1\}\to
\{0,\dots ,M-1\}$ where $N,\,M$ are not  powers of two
(then we extend them to the closest power of two). For more
details, see the next section.

\smallskip

Notice that the choice of gate array (Boolean circuit)
as the classical
model of computation is essential in the following sense:
a quantum routine cannot use conditional
instructions. Indeed, to implement such an
instruction we must observe the memory
in the midst of calculation, but the observation generally will
change its current quantum state.

\smallskip

In the same vein, we must avoid copying instructions, 
because the classical
copying operator $|x\rangle \to |x\rangle\otimes |x\rangle$
is not linear. In particular, each output qubit from a quantum gate
can be used only in one gate at the next step (if several
gates are used parallelly): cloning is not
allowed.

\smallskip

These examples show that the basics of quantum code 
writing will have a very distinct flavor.

\medskip

We now pass to the problems posed by the input/output
routines.  
 
\smallskip

Input, or initialization, in principle can be implemented
in the same way as a computation: we produce
an input state starting e.g. from the classical
state $|0\rangle$ and applying a sequence of basic
unitary operators: see the next section. Output, however, involves
an additional quantum mechanical notion: that of
{\it observation.}

\medskip

(E) The simplest model of observation of a quantum system
with the Hilbert space $H$ involves the choice of
an orthonormal basis of $H.$ Only elements of this
basis $|\chi_i\rangle$ can appear as the results
of observation. If our system is in some state
$|\psi\rangle$ at the moment of observation,
it will be observed in the state $|\chi_i\rangle$
with probability $|\langle \chi_i|\,\psi\rangle |^2.$

\smallskip

This means first of all that every quantum computation
is inherently probabilistic. Observing (a part of) the quantum memory
is not exactly the same as ``printing the output''.
We must plan a series of runs of the same quantum
program and the subsequent classical processing
of the observed results, and we can hope only to get
the desired answer with probability close to one.

\smallskip

Furthermore, this means that by implementing quantum
parallelism simplemindedly  as in (14), and then observing
the memory as if it were the classical $n$--bit register,
we will simply get some value $F(x)$ with
probability $1/N$. This does not use the potential  
of the quantum parallelism. Therefore we formulate a corrected version
of this notion, leaving more flexibility and stressing the
additional tasks of the designer, each of which eventually contributes
to the  complexity estimate.

\medskip

\proclaim{\quad 2.3. Quantum parallel processing: version II}
To solve efficiently a problem involving properties of the graph of
a function $F$, we must design:

\smallskip

(i) An auxiliary unitary operator $U$ 
carrying the relevant information about the graph of $F.$

\smallskip

(ii) A computationally feasible realization of $U$
with the help of standard quantum gates.

\smallskip

(iii) A computationally feasible realization
of the input subroutine.

\smallskip

(iv) A computationally feasible classical algorithm processing
the results of many runs of quantum computation.

\endproclaim 
 
\medskip

All of this must be supplemented by  quantum  
error--correcting encoding, which we will not
address here. In the next section we will discuss
some standard quantum subroutines.

\bigskip

\centerline{\bf 3. Selected quantum subroutines} 

\medskip

{\bf 3.1. Initialization.} Using the same conventions as in (14)
and the subsequent comments, in particular, the
identification $H_n=H_1^{\otimes n}$, we have
$$
\frac{1}{\sqrt{N}}\sum_{x=0}^{N-1} |x\rangle =
\frac{1}{\sqrt{N}}\sum_{\epsilon_i=0,1} |\epsilon_{n-1}\dots \epsilon_{0}\rangle =
\left(\frac{1}{\sqrt{2}}(|0\rangle +|1\rangle )\right)^{\otimes n} .
\eqno(15)
$$
In other words,
$$
\frac{1}{\sqrt{N}}\sum_{x=0}^{N-1} |x\rangle =
U_1^{(n-1)} \dots U_1^{(0)} |0 \dots 0\rangle
\eqno(16)
$$
where $U_1:\, H_1\to H_1$ is the unitary operator
$$
|0\rangle\mapsto \dfrac{1}{\sqrt{2}}\,(|0\rangle +|1\rangle),\
|1\rangle\mapsto \dfrac{1}{\sqrt{2}}\,(|0\rangle -|1\rangle) \, ,
$$
and $U_1^{(i)}=\roman{id}\otimes \dots \otimes U_1\otimes \dots 
\otimes \roman{id}$ acts only on the $i$--th qubit.

\smallskip

Thus making the quantum gate $U_1$ act on each memory
bit, one can in $n$ steps initialize our register in the state
which is the superposition of all $2^n$ classical states
with equal weights.

\medskip

{\bf 3.2. Quantum computations of classical functions.} Let $\Cal{B}$ be
a finite basis of classical gates containing one--bit
identity and generating all
Boolean circuits, and $F:\,\bold{F}_2^m\to \bold{F}_2^n$ 
a function. We will describe how to turn
a Boolean circuit of length $L$ calculating $F$
into another Boolean circuit of comparable length
consisting only of reversible gates, and calculating a modified
function, which however contains all information
about the graph of $F.$ Reversibility means that each step
is a bijection (actually, an involution) 
and hence can be extended to a unitary
operator, that is, a quantum gate. For a gate $f,$ define $\widetilde{f}(|x,y\rangle)=|x,f(x)+y\rangle$ as in 2.2(D)
above.

\smallskip

{\bf 3.2.1. Claim.} {\it A Boolean circuit $\Cal{S}$ of length $L$ in the basis
$\Cal{B}$ can be processed into the reversible
Boolean circuit $\widetilde{\Cal{S}}$ of length
$O((L+m+n)^2)$  calculating
a permutation $H:\,\bold{F}_2^{m+n+L}\to \bold{F}_2^{m+n+L}$
with the following property:
$$
H(x,y,0)=(x,F(x)+y,0)=(\widetilde{F}(x,y),0).
$$
Here $x,y,z$ have sizes $m,n,L$ respectively. }

\smallskip

{\bf Proof.} We will understand $L$ here as the
sum of sizes of the outputs of all gates 
involved in the description of $\Cal{S}.$ 
We first replace in $\Cal{S}$ each gate $f$ by its
reversible counterpart $\widetilde{f}.$ This involves
inserting extra bits which we put side by side
into a new register of total length $L.$ The resulting subcircuit will
calculate a permutation $K:\,\bold{F}_2^{m+L}\to\bold{F}_2^{m+L}$ 
such that $K(x,0)=(F(x),G(x))$ for some function $G$ (garbage).

\smallskip

Now add to the memory one more register of size $n$
keeping the variable $y.$ Extend $K$ to the permutation
$\overline{K}:\, \bold{F}_2^{m+L+n}\to\bold{F}_2^{m+L+n}$
keeping $y$ intact: $\overline{K}:\, (x,0,y)\mapsto
(F(x),G(x),y).$  Clearly, $\overline{K}$ is calculated by
the same boolean circuit as $K$, but with extended register.

\smallskip

Extend this circuit by  the one adding the contents of the first
and the third register: $(F(x),G(x),y)\mapsto (F(x),G(x),F(x)+y).$
Finally, build the last extension which calculates $\bar{K}^{-1}$
and consists of reversed gates calculating $\overline{K}$
in reverse order. This clears the middle register
(scratchpad) and produces $(x,0,F(x)+y).$ The whole circuit requires
$O(L+m+n)$ gates if we allow the application of them 
to not necessarily neighboring bits. Otherwise
we must  insert gates for local permutations which will
replace this estimate by $O((L+m+n)^2).$

\medskip

{\bf 3.3. Fast Fourier transform.} Finding the least period of
a function of one  real variable can be done by
calculating its Fourier transforms and looking
at its maxima. The same strategy is applied by Shor
in his solution of the factorization problem.
We will show now that the discrete Fourier transform $\Phi_n$
is computationally easy (quantum polynomial time). We define $\Phi_n:\,H_n\to H_n$ by
$$
\Phi_n (|x\rangle )=\frac{1}{\sqrt{N}}\sum_{c=0}^{N-1} |c\rangle\, 
\roman{exp}\,(2\pi i cx/N)\,
\eqno(17)
$$
In fact, it is slightly easier to implement directly the operator
$$
\Phi_n^t (|x\rangle )=\frac{1}{\sqrt{N}}\sum_{c=0}^{N-1} |c^t\rangle \,
\roman{exp}\,(2\pi i cx/N)\,.
\eqno(18)
$$
where $c^t$ is $c$ read from the right to the left.
The effects of the bit reversal can be then 
compensated at a later stage without difficulty.

\smallskip

Let $U_2^{(kj)}:\,H_n\to H_n,\, k<j,$ be the quantum gate
which acts on the pair of the $k$--th and $j$--th qubits
in the following way: it multiplies
$|11\rangle$ by $\roman{exp}\,(i\pi/2^{j-k})$
and leaves the remaining classical states
$|00\rangle,|01\rangle,|10\rangle$ intact.

\smallskip

\proclaim{\quad 3.3.1. Lemma} We have
$$
\Phi_n^t=\prod_{k=0}^{n-1} \left( U_1^{(k)}\prod_{j=k+1}^{n-1}
U_2^{(kj)}\right) .
\eqno(19)
$$
\endproclaim

\smallskip

By our rules of the game, (19) has polynomial length
in the sense that it involves only $O(n^2)$ gates.
However, implementation of  $U_2^{(kj)}$ requires
controlling variable phase factors which tend to $1$
as $k-j$ grows.  Moreover, arbitrary pairs
of qubits must allow quantum mechanical
coupling so that for large $n$ the interaction
between qubits must be non--local.
The contribution of these complications to
the notion of complexity cannot be estimated without
going into the details of physical arrangement. Therefore
I will add a few words to this effect.

\smallskip

The implementation of quantum register suggested
in [CZ] consists of a collection of ions (charged atoms)
in a linear harmonic trap (optical cavity). Two of the
electronic states of each ion are denoted $|0\rangle$ and
$|1\rangle$ and represent a qubit. Laser pulses
transmitted to the cavity through the optical fibers
and controlled by the classical computer
are used to implement gates and read out. The Coulomb repulsion
keeps ions apart (spatial selectivity) which allows the preparation
of each
ion separately in any superposition of
$|0\rangle$ and $|1\rangle$ by timing the laser
pulse properly and preparing its phase carefully.
The same Coulomb repulsion allows for collective
excitations of the whole cluster whose quanta
are called phonons. Such excitations
are produced by laser pulses as well
under appropriate resonance conditions.
The resulting resonance selectivity combined with the
spatial selectivity implements a controlled  entanglement
of the ions that can be used in order to simulate
two and three bit gates. For a detailed and lucid mathematical
explanation, see [Ts], Lecture 8.

\smallskip

Another recent suggestion ([GeC]) is to use a single molecule
as a quantum register, representing qubits
by nuclear spins of individual atoms, and 
using interactions through chemical bonds
in order to perform multiple bit logic.
The classical technique of nuclear magnetic resonance
developed since the 1940's, which  allows one to work with
many molecules simultaneously, provides
the start up technology for this project.

\medskip

{\bf 3.4. Quantum search.} All the subroutines described up to
now boiled down to some identities
in the unitary groups involving products of not too
many operators acting on subspaces of small dimension.
They did not involve output subroutines and therefore did
not ``compute'' anything in the traditional sense of
the word. We will now describe the beautiful
quantum search algorithm due to L.~Grover
which produces a new identity of
this type, but also demonstrates the effect of 
observation 
and the way one can use quantum entanglement in order to
exploit the potential of quantum parallelism.

\smallskip

We will treat only the simplest version. Let
$F:\,\bold{F}_2^n\to \{0,1\}$ be a function
taking the value $1$ at exactly one point $x_0.$
We want to compute $x_0.$ We assume that
$F$ is computable in polynomial time, or else
that its values are given by an oracle.
Classical search  for $x_0$ requires on the average
about $N/2$ evaluations of $F$ where $N=2^n.$

\smallskip

In the quantum version, we will assume that
we have a quantum Boolean circuit (or quantum oracle)
calculating the unitary operator $H_n\to H_n$
$$
I_F:\, |x\rangle \mapsto e^{\pi iF(x)}|x\rangle .
$$
In other words, $I_F$ is the reflection inverting
the sign of $|x_0\rangle$ and leaving the remaining classical
states intact. 

\smallskip

Moreover, we put $J=-I_\delta$, where $\delta :\bold{F}_2^n\to
\{0,1\}$ takes 
the value $1$ only at $0,$ and $V=U_1^{(n-1)}\dots U_1^{(0)},$
as in (16).

\medskip

{\bf 3.4.1. Claim.} {\it (i) The real plane in $H_n$ 
spanned by 
the uniform superposition $\xi$ of all classical states (15)
and by $|x_0\rangle$
is invariant with respect to $T:=VJVI_F.$

\smallskip

(ii) $T$ restricted to this plane is the rotation 
(from $\xi$ to $|x_0\rangle$) by the angle $\varphi_N$ where}
$$
\roman{cos}\,\varphi_N =1-\frac{2}{N},\  \roman{sin}\,\varphi_N =  
2\,\frac{\sqrt{N-1}}{N}.
$$ 

\smallskip

The check is straightforward. 
 
\smallskip

Now, $\varphi_N$ is close to $\dfrac{2}{\sqrt{N}}$, and for the initial
angle $\varphi$ between $\xi$ and $|x_0\rangle$ we
have 
$$ 
\roman{cos}\,\varphi =-\frac{1}{\sqrt{N}}.
$$
Hence in $[\varphi /\varphi_N]\approx \dfrac{\pi\sqrt{N}}{4}$
applications of $T$ to $\xi$ we will get the state very close to $|x_0\rangle$.
Stopping the iteration of $T$ after as many steps
and measuring the outcome in the basis of classical states,
we will obtain $|x_0\rangle$ with probability very close to one.

\smallskip

One application of $T$ replaces in the quantum
search one evaluation of $F.$ Thus, thanks to quantum parallelism,
we achieve a polynomial speed--up in comparison
with the classical search. The case when $F$ takes value $1$
at several points and we only want to find one of them,
can be treated by an extension of this method. 
If there are $n$ such points, the algorithm requires
about $\sqrt{N/n}$ steps, and $n$ need not be known 
a priori: see [BoyBHT].

\bigskip

\centerline{\bf 4. Shor's factoring algorithm}

\medskip

{\bf 4.1. Notation.} Let $M$ be a number to be factored. 
We will assume that it is odd
and is not a power of a prime number.

\smallskip

Denote by $N$ the size of the basic memory register
we will be using (not counting scratchpad). Its bit size $n$
will be about twice that of $M$. More precisely,
choose $M^2<N=2^n<2M^2.$ Finally, let $1<t<M$ be a
random parameter with $\roman{gcd}\,(t,M)=1.$
This condition can be checked classically
in time polynomial in $n.$

\smallskip

Below we will describe one run of Shor's algorithm,
in which $t$ (and of course, $M$, $N$) is fixed.
Generally, polynomially many
runs will be required, in which the value of $t$ 
can remain the same or be chosen anew. This is needed 
in order to gather statistics. Shor's algorithm
is a probabilistic one, with two sources of randomness
that must be clearly distinguished. One is built into
the classical probabilistic reduction of factoring to the finding
of the period of a function. Another stems from the
necessity of observing quantum memory, which, too,
produces random results.

\smallskip

More precise estimates than those given here show that
a quantum computer which can store about $3n$ qubits
can find a factor of $M$ in time of order $n^3$ 
with probability close to $1:$ see [BCDP].
On the other hand, it is widely believed
that no recursive function of the type $M\mapsto$
{\it a proper factor of} $M$ belongs to $PF.$
This is why the most popular public key encryption schemes
rely upon the difficulty of the factoring problem.

\medskip

{\bf 4.2. Classical algorithm.} Put 
$$
r:=\roman{min}\,\{\rho\,|\,t^{\rho}\equiv 1\,\roman{mod}\,M\}
$$
which is the least period of $F:\,a\mapsto t^a\,\roman{mod}\,M.$

\smallskip

{\bf 4.2.1. Claim.} {\it If one can efficiently calculate $r$
as a function of $t,$
one can find a proper divisor of $M$ in polynomial in $\roman{log}_2M$
time with probability $\ge 1-M^{-m}$ for any fixed $m.$}

\smallskip

Assume that for a given $t$ the period $r$ satisfies
$$
r\equiv 0\,\roman{mod}\,2,\ t^{r/2}\neq -1\,\roman{mod}\,M
$$
Then $\roman{gcd}\,(t^{r/2}+1,M)$
is a proper divisor of $M.$ Notice that $\roman{gcd}$
is computable in polynomial time.

\smallskip

The probability that this
condition holds is $\ge 1-\dfrac{1}{2^{k-1}}$ where $k$ is the number
of different odd prime divisors of $M$, hence $\ge \dfrac{1}{2}$
in our case.
Therefore we will find a good $t$ with probability
$\ge 1-M^{-m}$ in   $O(\roman{log}\,M)$ tries.
The longest calculation in one try is that of
$t^{r/2}.$ The usual squaring
method takes polynomial time as well.

\medskip

{\bf 4.3. Quantum algorithm calculating $r$.} Here we describe
one run of the quantum algorithm which purports to
compute $r$, given $M,N,t.$ We will use the working
register that can keep a pair consisting of a variable $0\le a\le N-1$
and the respective value of the function $t^a\,\roman{mod}\,M.$
One more register will serve as the scratchpad
needed to compute $|a, t^a\,\roman{mod}\,M\rangle$
reversibly. When this calculation is completed,
the content of the scratchpad will be reversibly erased: cf. 3.2.1.
In the remaining part of the computation the scratchpad will
not be used anymore, we can decouple it, and forget about it.

\smallskip

The quantum computation consists of four steps, three of which were described 
in sec. 3:

\smallskip

(i) Partial initialization produces from $|0,0\rangle$ the superposition
$$
\frac{1}{\sqrt{N}}\sum_{a=0}^{N-1}|a,0\rangle.
$$

\smallskip

(ii) Reversible calculation of $F$ processes this state into
$$
\frac{1}{\sqrt{N}}\sum_{a=0}^{N-1}|a, t^a\,\roman{mod}\,M\rangle.
$$

\smallskip

(iii) Partial Fourier transform then furnishes 
$$
\frac{1}{N}\sum_{a=0}^{N-1}\sum_{c=0}^{N-1}
\roman{exp}\,(2\pi iac/N)\,|c, t^a\,\roman{mod}\,M\rangle.
$$

\smallskip

(iv) The last step is the observation of this state with respect
to the system of classical states $|c, m\,\roman{mod}\,M\rangle.$
This step produces some concrete output
$$
|c, t^k\,\roman{mod}\,M\rangle
\eqno(20)
$$
with probability
$$
\left| \frac{1}{N}\sum_{a:\,t^a\equiv t^k\,\roman{mod}\,M}
\roman{exp}\,(2\pi iac/N)\right|^2.
\eqno(21)
$$
The remaining part of the run is assigned to the classical computer
and consists of the following steps.

\smallskip

(A) {\it Find the best approximation (in lowest terms)
to $\dfrac{c}{N}$ with 
denominator $r^{\prime}<M<\sqrt{N}$:}
$$
\left|\frac{c}{N} -\frac{d^{\prime}}{r^{\prime}}\right|<\frac{1}{2N}.
\eqno(22)
$$
\smallskip

As we will see below, we may hope that $r^{\prime}$ will
coincide with $r$ in at least one run among at most polynomially
many. Hence we try $r^{\prime}$ in the role of $r$ right away:

\smallskip

(B) {\it If $r^{\prime}\equiv 0\,\roman{mod}\,2$, calculate
$\roman{gcd}\,(t^{r^{\prime}/2}\pm 1,M).$}

\smallskip

If $r^{\prime}$ is odd, or if $r^{\prime}$ is even, but
we did not get a proper divisor of $M$, repeat the
run $O(\roman{log}\,\roman{log}\,M)$ times with the same $t.$ 
In case of failure, change $t$ and start a new run.

\medskip

{\bf 4.3.1. Justification.} We will now show that, given $t,$
from the observed values of $|c, t^k\,\roman{mod}\,M\rangle$
in $O(\roman{log}\,\roman{log}\,M)$ runs we can find the correct value
of $r$ with probability close to $1.$

\smallskip

Let us call the observed value of $c$ {\it good}, if
$$
\exists\, l\in\left[-\frac{r}{2},\frac{r}{2}\right],\ rc\equiv l\,\roman{mod}\,N.
$$

\smallskip

In this case there exists such $d$ that 
$$
-\frac{r}{2}\le rc-dN=l\le \frac{r}{2}
$$
so that
$$
\left|\frac{c}{N} -\frac{d}{r}\right|<\frac{1}{2N}.
$$
Hence if $c$ is good, then $r^{\prime}$ found from (22)
in fact divides $r.$

\smallskip

Now call $c$ {\it very good} if $r^{\prime}=r.$

\smallskip

Estimating the exponential sum (21), we can easily check
that the probability of observing a good $c$
is $\ge \dfrac{1}{3r^2}.$ On the other hand, there are $r\varphi (r)$
states $|c,t^k\,\roman{mod}\,M\rangle$ with very good $c.$ 
Thus to find a very good $c$
with high probability, $O(r^2\,\roman{log}\,r)$ runs
will suffice.

\newpage

\centerline{\bf 5. Kolmogorov complexity and growth of recursive functions}

\medskip

Consider general functions $f:\,\bold{N}\to \bold{N}.$ 
Computability theory uses several growth scales
for such functions, of which two are most useful:
$f$ may be majorized by some recursive function
(e.g. when it is itself recursive), or by a polynomial
(e.g. when it is computable in polynomial time).
Linear growth does not seem particularly
relevant in this context. However, this impression is
quite misleading, at least if one allows re--ordering $\bold{N}.$
In fact, we have:

\medskip

{\bf 5.1. Claim.} {\it There exists a permutation
$\bold{K}:\,\bold{N}\to\bold{N}$ such that for any partially
recursive function $f:\,\bold{N}\to \bold{N}$
there exists a constant $c$ with the property
$$
\bold{K}\circ f\circ \bold{K}^{-1}(n)\le c\,n\ \roman{for\ all}\ n\in 
\bold{K}(D(f)).
\eqno(23)
$$
Moreover, $\bold{K}$ is bounded by a linear function, but
$\bold{K}^{-1}$ is not bounded by any recursive function.}

\smallskip

{\bf Proof.} We will use the Kolmogorov complexity measure.
 For a recursive function
$u:\,\bold{N}\to\bold{N},\, x\in\bold{N},$ put $C_u(x):=\roman{min}\,
\{k\,|\,f(k)=x\},$ or $\infty$ if such $k$ does not exist.
Call such a function $u$ {\it optimal} if, for any other recursive
function $v,$ there exists a constant $c_{u,v}$ such that
$C_u(x)\le c_{u,v}C_v(x)$ for all $x.$  Optimal functions do exist
(see e.g. [Ma1], Theorem VI.9.2);
in particular, they take all positive integer values
(however they certainly are not everywhere defined). Fix one such $u$
and call $C_u(x)$ the (exponential) complexity of $x.$
By definition, $\bold{K}=\bold{K}_u$ rearranges $\bold{N}$
in the order of increasing complexity. In other words,
$$ 
\bold{K}(x):=1+\roman{card}\,\{y\,|\,C_u(y)<C_u(x)\}.
\eqno(24)
$$
We first show that 
$$
\bold{K}(x)=\roman{exp}\,(O(1))\,C_u(x).
\eqno(25)
$$ 
Since $C_u$ takes each value at most once, it follows from
(24) that $\bold{K}(n)\le C_u(n).$ In order to show that
$C_u(x)\le c\,\bold{K}(x)$ for some $c$ it suffices to check
that 
$$
\roman{card}\,\{k\le N\,|\,\exists\,x,\,C_u(x)=k\}\ge b\,N
$$
with some $b>0.$ In fact, at least half of the numbers $x\le N$ have
the complexity which is no less than $x/2.$ 

\smallskip

Now, VI.9.7(b) in [Ma1] implies that, for any recursive function
$f$ and all $x\in D(f),$ we have $C_u(f(x))\le \roman{const}\,C_u(x).$ 
Since $C_u(x)$ and $\bold{K}(x)$ have the same order of growth
up to a bounded factor, our claim follows.

\medskip

{\bf 5.2. Corollary.} {\it Denote by $S_{\infty}^{\roman{rec}}$
be the group of recursive permutations of $\bold{N}.$
Then $\bold{K}\,S_{\infty}^{\roman{rec}}\,\bold{K}^{-1}$
is a subgroup of permutations of no more than linear growth.}

\medskip

Actually, appealing to the
Proposition VI.9.6 of [Ma1], one can considerably
strengthen this result. For example, let $\sigma$ be a recursive permutation, $\sigma^{\bold{K}}=
\bold{K}\sigma {\bold{K}}^{-1}.$ Then $\sigma^{\bold{K}}(x)\le c x$
so that $(\sigma^{\bold{K}})^n(x)\le c^n x$ for $n>0.$ But actually the last
inequality can be replaced by 
$$
(\sigma^{\bold{K}})^n(x)\le c^{\prime}n
$$
for a fixed $x$ and variable $n$. With both $x$ and $n$ variable
one gets the estimate $O(xn\,\roman{log}\,(xn))$.

\medskip

In the same way as finite permutations appear in the quantum versions
of Boolean circuits, infinite (computable) permutations
are natural for treating quantum Turing machines ([Deu])
and our normal computation models. In fact,
if one assumes that the  transition function $s$ is a permutation,
and then extends it to the unitary operator $U_s$
in the infinite--dimensional Hilbert space, one
might be interested in studying the spectral properties of such operators.
But the latter  depend only on the conjugacy class. 
Perhaps the universal
conjugation $U_{\bold{K}}$ might be a useful theoretical
tool in this context. In the purely classical situation,
(23) may play a role in studying the limiting behavior
of polynomial time algorithms, as suggested  in [Fr1] and [Fr2].

\medskip

Finally, I would like to comment upon the hidden role
of Kolmogorov complexity in the real life of
classical computing. The point is that in a
sense (which is difficult to formalize), we
are interested only in the calculation of sufficiently
nice functions, because a random Boolean
function will have (super)exponential complexity anyway.
A nice function, at the very least, has a short
description and, therefore, a small Kolmogorov
complexity. Thus,  dealing with
practical problems, we actually work not with small
numbers, graphs, circuits, $\dots$ , but rather with an
initial segment of the respective constructive
world reordered with the help of $\bold{K}.$
We systematically replace a large object by its short description,
and then try to overcome the computational difficulties
generated by this replacement.

\bigskip

\centerline{\bf Appendix}

\medskip

The following text is a contribution to the prehistory
of quantum computing.
It is the translation from Russian of the last three
paragraphs of the Introduction to [Ma2] (1980).
For this reference I am grateful to A.~Kitaev [Ki].

\medskip

`` Perhaps, for better understanding
of this phenomenon [DNA replication], we need
a mathematical theory of quantum automata.
Such a theory would provide us with mathematical models of
deterministic processes with quite unusual properties.
One reason for this is that the quantum state space
has far greater capacity than the classical one:
for a classical system with $N$ states, its quantum
version allowing superposition accommodates $c^N$ 
states. When we join two classical systems,
their number of states $N_1$ and $N_2$ are multiplied,
and in the quantum case we get  exponential growth
$c^{N_1N_2}.$

\smallskip

These crude estimates show that the quantum behavior 
of the system
might be much more complex than its classical simulation.
In particular, since there is no unique decomposition
of a quantum system into its constituent parts,
a state of the quantum automaton can be considered in many ways
as a state of various virtual classical automata.
Cf. the following instructive comment at the end of
the article [Po]: `The quantum--mechanical computation 
of one molecule of methane requires $10^{42}$ grid points.
Assuming that at each point we have to perform only
10 elementary operations, and that the computation
is performed at the extremely low temperature
$T=3.10^{-3}K,$ we would still have to use all the
energy produced on Earth during the last century.'

\smallskip

The first difficulty we must overcome is the choice
of the correct balance between the mathematical and the physical
principles. The quantum automaton has to
be an abstract one: its mathematical model must appeal
only to the general principles of quantum physics, without
prescribing a physical implementation. Then the model
of evolution is the unitary rotation in a 
finite dimensional Hilbert space, and the decomposition
of the system into its virtual parts corresponds
to the tensor product decomposition of the state space.
Somewhere in this picture we must accommodate interaction,
which is described
by density matrices and probabilities.''

\bigskip

\centerline{\bf Bibliography}

\medskip

[BCDP] D.~Beckman, A.~N.~Chari, Sr.~Devabhaktuni, J.~Preskill.
{\it Efficient networks for quantum computing.} Phys. Rev. A,
54:2 (1996), 1034--1063.

\smallskip

[Ben1] P.~Benioff. {\it The computer as a physical system:
A microscopic quantum mechanical Hamiltonian model of computers
as represented by Turing machines.} J.~Stat.~Phys., 22 (1980), 563--591.

\smallskip

[Ben2] P.~Benioff. {\it Quantum mechanical Hamiltonian models 
of Turing machines that dissipate no energy.} Phys.~Rev.~Lett.,
48 (1980), 1581--1585.

\smallskip

[BoL] D.~Boneh, R.~Lipton. {\it Quantum cryptoanalysis
of hidden linear functions.} Proc. of Advances in Cryptology
--- CRYPTO '95, Springer LN in Computer Science,
vol. 963 (1995), 424--437.

\smallskip

[BoyBHT] M.~Boyer, G.~Brassard, P.~H\o yer, A.~Tapp.
{\it Tight bounds on quantum searching.}
Preprint, 1996.

\smallskip

[CZ] J.~Cirac, P.~Zoller. {\it Quantum computation with cold
trapped ions.} Phys. Rev. Lett., 74:20 (1995), 4091--4094.

\smallskip

[Deu] D.~Deutsch. {\it Quantum theory, the Church--Turing principle
and the universal quantum computer.} Proc.~R.~Soc.~Lond.
A 400 (1985), 97--117.

\smallskip

[DeuJ] D.~Deutsch, R.~Jozsa. {\it Rapid solutions of problems
by quantum computation.} Proc.~Roy.~Soc. London, Ser. A, 449 (1992), 553--558.

\smallskip

[Fe1] R.~Feynman. {\it Simulating physics with computers.} Int. J.
of Theor. Phys., 21 (1982), 467--488.

\smallskip

[Fe2] R.~Feynman. {\it Quantum mechanical computers.}
Found. Phys., 16 (1986), 507--531.

\smallskip

[Fr1] M.~Freedman. {\it Topological views on
computational complexity.} In Proc. ICM Berlin 1998,
vol. II, 453--464.

\smallskip

[Fr2] M.~Freedman. {\it Limit, logic, and computation.}
Proc. Nat. Ac. Sci. USA, 95 (1998), 95--97.

\smallskip

[Fr3] M.~Freedman. {\it P/NP, and the quantum field computer.}
Proc. Nat. Ac. Sci. USA, 95 (1998), 98--101.

\smallskip

[GaJ] M.~Garey, D.~Johnson. {\it Computers and Intractability:
A Guide to the Theory of NP--Completeness.} W.~H.~Freeman and Co.,
San--Francisco, 1979.

\smallskip

[GeC] N.~Gershenfield, I.~Chuang. {\it Bulk spin--resonance quantum
computation.} Science 275 (1997), 350--355.

\smallskip

[Gri] D.~Grigoriev. {\it Testing the shift--equivalence of polynomials
using quantum mechanics.} In: Manin's Festschrift, Journ. of Math. Sci.,
82:1 (1996), 3184--3193.

\smallskip

[Gro] L.~K.~Grover. {\it Quantum mechanics helps in searching for a needle
in a haystack.} Phys.~Rev.~Lett. 79 (1997), 325--328.

\smallskip

[Ki1] A.~Kitaev. {\it Quantum computations: algorithms and
error correction.} Russian Math. Surveys, 52:6 (1997), 53--112.

\smallskip

[Ki2] A.~Kitaev. {\it Classical and quantum computations.}
Lecture notes, Independent University, Moscow, 1998.

\smallskip

[Ma1] Yu.~Manin. {\it A Course in Mathematical Logic.}
Springer Verlag, 1977, pp. xiii+286.

\smallskip

[Ma2] Yu.~Manin. {\it Computable and uncomputable (in Russian).}
Moscow, Sovetskoye Radio, 1980.

\smallskip

[Mu] D.~Mumford. {\it The statistical description of visual signals.}
Preprint.

\smallskip

[Po] R.~P.~Poplavskii. {\it Thermodynamical models of
information processing (in Russian).} Uspekhi Fizicheskikh Nauk,
115:3 (1975), 465--501.

\smallskip
[Sa] A.~Salomaa. {\it Computation and Automata.} Cambridge UP, 1985.

\smallskip

[Sh] P.~W.~Shor. {\it Polynomial--time algorithms for prime
factorization and discrete logarithms on a quantum computer.}
SIAM J. Comput., 26:5 (1997), 1484--1509.

\smallskip

[Si] D.~Simon. {\it On the power of quantum computation.}
Proc. of the 35th Ann. Symp. on Foundations of Comp. Sci.
(1994), 116--123.

\smallskip

[Ts] B.~Tsirelson. {\it Quantum information processing.}
Lecture notes, Tel--Aviv University, 1997.

\enddocument